\documentclass[twocolumn,prc,aps,superscriptaddress,showpacs,floatfix]
{revtex4}

\def\be{\begin{eqnarray}}
\def\ee{\end{eqnarray}}
\def\bea{\begin{eqnarray}}
\def\eea{\end{eqnarray}}
\def\beas{\begin{eqnarray*}}
\def\eeas{\end{eqnarray*}}

\usepackage{graphics}
\usepackage{verbatim}
\usepackage{times}
\usepackage{epsfig}
\usepackage{eurosym}
\usepackage{array}
\begin{document}

\bibliographystyle{apsrev}

\title{3D Printing of Scintillating Materials}

\author{Y. Mishnayot}
\affiliation{Racah Institute of Physics, Hebrew University of Jerusalem, Jerusalem, Israel 91904}

\author{M. Layani}
\affiliation{Casali Center,Institute of Chemistry and the Center for Nanoscience and Nanotechnology, The Hebrew University of Jerusalem, Jerusalem, Israel 91904}

\author{I. Cooperstein}
\affiliation{Casali Center,Institute of Chemistry and the Center for Nanoscience and Nanotechnology, The Hebrew University of Jerusalem, Jerusalem, Israel 91904}

\author{S. Magdassi}
\affiliation{Casali Center,Institute of Chemistry and the Center for Nanoscience and Nanotechnology, The Hebrew University of Jerusalem, Jerusalem, Israel 91904}

\author{G. Ron}
\email{gron@phys.huji.ac.il}
\affiliation{Racah Institute of Physics, Hebrew University of Jerusalem, Jerusalem, Israel 91904}

\date{\today}
\begin{abstract}
We demonstrate, for the first time, the applicability of 3D printing technique to the manufacture 
of scintillation detectors. We report of a formulation, usable in stereolithographic printing, that 
exhibits scintillation efficiency on the order of 30\% of that of commercial polystyrene based scintillators
We discuss the applicability of these techniques and propose future enhancements that will allow
tailoring the printed scintillation detectors to various application.
\end{abstract}

\maketitle


\section{\label{sec:Intro}Introduction}
Scintillating materials are commonly used in charged particle detectors as detector elements due to their ease of 
manufacture, relatively low cost, and good timing resolution. The most common type of scintillator is the plastic scintillator, in
which wavelength shifting dopants are embedded in a polystyrene matrix. A charged particle, traversing the polystyrene base
excited some of the polymer molecules, causing them to emit UV light upon recombination. Typically, 2,5-diphenyloxazole
(PPO) is used as a first step to down-convert the UV light to longer wavelength, peaking at 350nm. Due to the short
absorption length of the polymer base in the UV range, a second wavelength shifting dopant, typically 
1,4-di-(5-phenyl-2-oxazolyl)-benzene (POPOP), with an emission wavelength peak at 410 nm is used to further 
shift the emitted light to the visible range, while keeping the wavelength short enough to be effectively detected by the 
detection element (usually a photocathode).

Scintillators are nowadays typically made by either casting of a resin+hardener combination, or by extrusion of a molten
scintillator. These cast or extruded scintillator bar are then machined to the required geometry using standard 
machining techniques (note that the low melting/softening point of scintillators and to the requirements of a 
clean surface for light collection places severe constraints on the requirements from the machine shop, 
i.e., low temperature machining and extreme cleanliness with the additional need to polish the scintillator 
parts after machining). We note that while cast scintillators are more versatile in their possible designs, 
they suffer from long production times (up to 2 weeks for polymerization) and are also not amenable to complex 
geometries due to the requirement of producing a complicated mold.

We approach the problem of scintillator manufacturing differently. Rather than taking a 
top-down approach (i.e. machining a solid block into a preferred shape Ð subtractive manufacturing), we take 
a bottom-up approach, using newly developed 3D printing techniques to directly print the required form 
(additive manufacturing), also allowing the creation of scintillator designs that cannot be achieved 
using a standard approach (e.g., hollow, gas filled, scintillators, or scintillator designs with features with are 
too small to be machined, both directly onto the scintillator or as a mold).

Three dimensional printing has evolved into a paradigm shifting technology in recent years. The ability to 
replicate and design, at home, structures down to the 10s of microns scale at less than 10 K\$ is at hand. 
Several technologies have emerged, including laser sintering, Fused Deposition Modeling (FDM via extrusion), 
and photopolymerization. FDM 3D printing is performed by the extrusion of a plastic filament through a thin 
nozzle, constructing a geometry layer by layer. Photopolymerization (PP) approaches use UV light to 
selectively polymerize a liquid, either inside a resin bath, or in a sprayed layer, similar to the method used in 
inkjet printers~\cite{Print1,Print2}. 

\section{Experiment}
In order to test the applicability of additive manufacturing techniques to the manufacture of scintillators 
we designed a formulations, which is UV-polymerizable, based on an acrylic monomer, and doped with 
different fractions of scintillating and wavelength shifting materials. We roughly follow the recipes provided
by~\cite{Scint1} since they have been shown to achieve efficiencies on the order of several 10s of percent.

Ink preparation: The curable formulation was composed of 99.5 wt\%  SR9035 (Sartomer, Arkema) as the monomer 
and 0.5 wt\%  Lucirin TPO (BASF) as the photoinitiator. The formulation was stirred for 30 minutes in a water bath at 60~$^\circ$C. 
To this formulation the following scintlator components were added at various concentrations (see Tab.~\ref{tab:Res}) 2,5-Diphenyloxazole, (PPO, Sigma-Aldrich), 
1,4-Bis(5-phenyl-2-oxazolyl)benzene (POPOP, Sigma-Aldrich), and Naphtalene  (Sigma-Aldrich). The final formulation was stirred for 1 hour in a water bath at 60 $^\circ$C.

\begin{figure}[ht]
\includegraphics[width=0.4\textwidth]{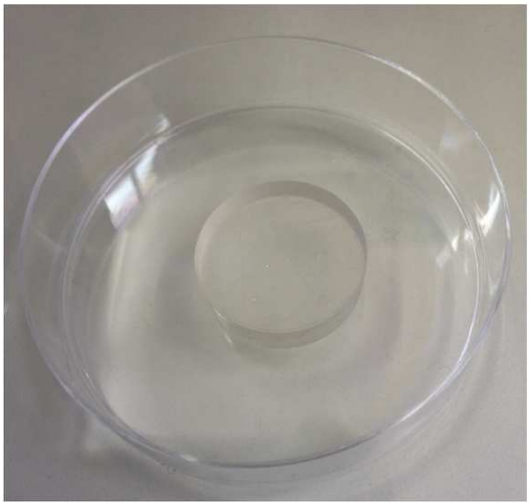}
\caption{\label{fig:Samp} (Color online) A picture of the printed output of one of our representative samples.}
\end{figure}

The compounds were polymerized into the required shape (a cylinder 6.3mm in height and 20mm diameter) 
using an Asiga Pico Plus 39 printer (Asiga, Australia). 2 different layer thicknesses were tested, 25 $\mu$m, and 127 $\mu$m, in order to test the effect of the interface 
between printed layers on the light collection efficiency. The printed output was compared to reference samples of clear PMMA (plexiglas) and EJ-204 scintillator 
(Eljen Ltd.) of the same dimensions. Fig.~\ref{fig:Samp} is a photograph of the output of one of our printed samples containing 
the scintillating compounds.

\begin{figure}[ht]
\includegraphics[width=0.5\textwidth]{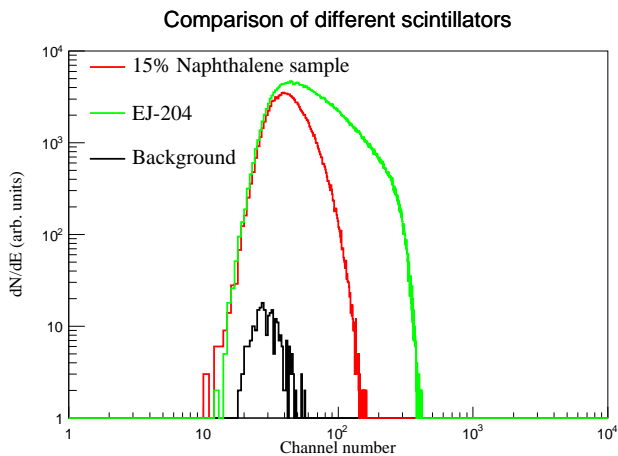}
\caption{\label{fig:Hist} (Color online) Typical spectrum from a reference sample (EJ-204), our printed sample, and the background measurement.
The printed sample achieved $\sim$ 30\% efficiency compared to commercial scintillator material, proving the viability of this technique.
 }
\end{figure}

All samples were tested in a light tight box using both background cosmic radiation
and a 15.1 KBq $^{90}$Sr source. The samples were matched to a photomultiplier (Hamamatsu R647)
operated at 1 kV, through a 100mm Perspex light guide. Light output was readout from the PMT into 
a CAEN charge integrating digitizer (V1720) and histograms were recorded to a PC. A typical histogram from the EJ-204 sample and our best sample is shown in Fig.~\ref{fig:Hist}.

\begin{figure}[ht]
\includegraphics[width=0.5\textwidth]{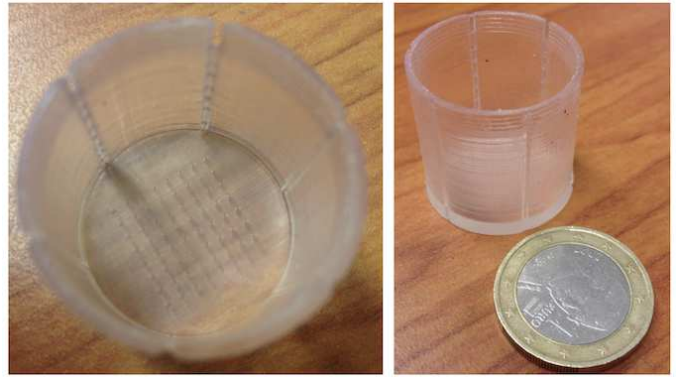}
\caption{\label{fig:ActiveTarget} (Color online) Two views of the Active Target prototype (see text), note
the external grooves, meshed cap, and threaded area. \EUR{1} coin included for scale.
 }
\end{figure}

In addition, several prints were made of scintillators with more complex geometries, in order to show that such
geometries may be printed using our formulated materials. Since this geometries are almost impossible to machine, 
a comparison to a standard sample was not made. In particular, a geometry relevant for the Frozen Spin Active 
Target~\cite{ActiveTarget}
at the Crystal Ball detector at MAMI~\cite{CB} was printed, consisting of a thin walled, hollow cylinder, capped with a mesh to 
allow for the flow of super-fluid helium, and including external grooves for coupling to wave length shifting fibers.
Fig.~\ref{fig:ActiveTarget} shows the printed design, note the mesh and the threaded top of the cylinder, which is almost impossible
to create in a scintillator using standard production techniques.

\section{Results}
\begin{figure}[ht]
\includegraphics[width=0.5\textwidth]{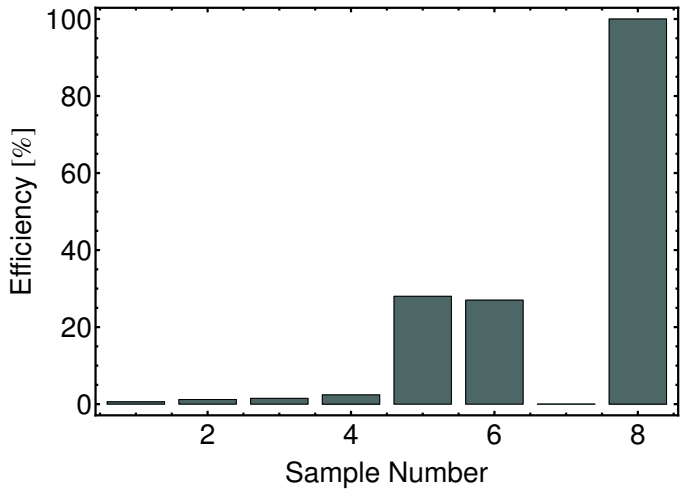}
\caption{\label{fig:Eff} Efficiency measured for the different samples listed in Tab.~\ref{tab:Res}, compared to 
a commercial scintillator (sample 8).}
 \end{figure}

Several different formulations were tested, with varying concentrations of the activator (Naphtalene) and the wavelength shifters. 
Table~\ref{tab:Res} summarizes the results of the different formulations, compared to a reference sample (EJ-204). The table
clearly shows that efficiencies on the order of 30\% are easily achievable using 3D printing, demonstrating the viability 
of such techniques. It should be noted that acrylic based scintillator are known to be less efficient that aromatic scintillators (such as 
our reference sample) and we attribute part of the efficiency loss to this difference, the other part being attributed to both 
optical quality of the printed sample and possible degradation of the active materials in the printed sample during the UV polymerization phase. Fig.~\ref{fig:Eff} summarizes the efficiency achieved for the various samples listed in Tab.~\ref{tab:Res}.

\begin{table*}[htp]
\begin{center}
 \begin{tabular}{|c|c|c|c|c|c|c|} \hline
  Number & Layer thickness ($\mu$m)& Naphthalene content (\%) & PPO content (\%) & POPOP content (\%) & Efficiency (\%)& Comments\\ \hline \hline
  1 & 127 & - & 1 & 0.05 & 0.6& - \\ \hline
  2 & 127 & 3 & 1 & 0.05 & 1.2&- \\ \hline
  3 & 127 & 3 & 1 & 0.05 & 1.5& - \\ \hline
  4 & 25 & 3 & 1 & 0.05 & 2.4& - \\ \hline
  5 & 25 & 15 & 1.5 & 0.08 & 28& - \\ \hline
  6 & 127 & 15 & 1.5 & 0.08 & 27&- \\ \hline
  7 & - & - & - & - & 0&Clear Sample \\ \hline
  8 & - & - & - & - & 100&EJ-204 Scintillator \\ \hline
 \end{tabular} 
 \caption{\label{tab:Res}Results of the comparison between the different formulations tested.}
\end{center}
\end{table*}

Additionally, we have seen evidence of degradation with time of the scintillation efficiency. We attribute this degradation to the sublimation 
of the Naphtalene activator at high concentrations. In the future modifications of the formulation we use will investigate the replacement of the 
base polymer by a different polymer, or replacement of the Naphtalene activator by a different, less volatile compound (for example, 1,1,3-Trimethyl-3-phenylindane (TMPI),
proposed in~\cite{Scint1}).
\section{Conclusions}
We have shown, for the first time, that additive manufacturing techniques are a viable alternative
for scintillator manufacturing. In particular, such techniques are well suited for the production of small, complex,
designs, which are hard to produce using either extrusion or machining techniques. While our current system
uses a standard scintillating compound, designed for the detection of charged particles, a logical extension of our method
is the inclusion of additional dopants, such as for example, high-Z materials, or Gd, which can enhance detection capabilities 
for neutral particles such as photons and neutrons respectively. This doping can be achieved by the simple technique of mixing 
particles of the required dopants with our formulation prior to printing. An additional possible extension is the 
introduction of methods which are capable of printing using multiple compounds simultaneously, or varying the 
concentration of dopants in a continuos manner (such as InkJet based printers). These further improvements will
allow, for example, for the manufacturing of scintillation detectors with embedded wavelength shifting fibers, removing the 
for need coupling using optical grease.

\section{Acknowledgements}
This work was supported by the Ernest Bergman Fund. 
The authors would like to acknowledge Dr. A. Thomas from the Institute for Nuclear Physics at Mainz, and Prof. 
E. Downie, from the George Washington University, for helpful discussion regarding the A2 Active Target.

\end{document}